\documentclass[preprint,12pt]{elsarticle}
\usepackage{graphicx}
\newcommand{\beq}{\begin{equation}}
\newcommand{\eeq}{\end{equation}}
\newcommand{\ba}{\begin{array}}
\newcommand{\ea}{\end{array}}
\newcommand{\beqa}{\begin{eqnarray}}
\newcommand{\eeqa}{\end{eqnarray}}
\newcommand{\bd}[1]{ \mbox{\boldmath $#1$}  }

\newcommand{\lam}{\lambda}

\newcommand{\JPG}[1]{J.Phys.{\bf G{#1}}}
\newcommand{\NPA}[1]{Nucl.Phys.{\bf A{#1}}}

\newcommand{\PRC}[1]{Phys.Rev.{\bf C{#1}}}

\newcommand{\PR}[1]{Phys.Rep.{\bf {#1}}}

\begin{document}


\title{POPULATION OF ROTATIONAL STATES IN THE GROUND-BAND 
OF FISSION FRAGMENTS}
\author{{\c S}erban Mi\c sicu}
\address{Department of Theoretical Physics, National Institute for Physics and Nuclear Engineering 
Horia Hulubei, Atomistilor 407, RO-077125, POB-MG6, M\u{a}gurele-Bucharest, Romania, EU\\Email:{\em misicu@theory.nipne.ro}}


\begin{abstract}
The population of rotational states in the ground-state band of neutron-rich 
fragments emitted in the spontaneous fission of $^{252}$Cf is described within 
a time-dependent quantum model similar to the one used for Coulomb excitation. 
The initial population probability of the states included in the selected basis 
is calculated according to the bending model at scission. Subsequently 
these initial amplitudes are feeding the coupled dynamical equations 
describing the population of rotational states in both fragments during the 
tunneling and post-barrier (pure Coulomb) motion. As application we consider the 
high yield Mo-Ba pair for different number of emitted neutrons.
\end{abstract}

\maketitle


\section{Introduction}\label{aba:sec1}

We owe much to the chief cultivator of nuclear structure models in Romania, Prof.Dr.A.A. Raduta,
whose oustanding work as a researcher and teacher during the last four decays contributed essentially
to the formation of an authentic school of nuclear theory at M\u agurele. I was initiated as a researcher
in nuclear theory under his close advisorship starting with 1988. His commitment to high quality 
research was a strong example and inspiration for me and contributed descisively in shaping my scientific
career. Among the multiple themes addressed by Prof.Dr.A.A. Raduta in nuclear structure physics, he payed a special attention to the development of phenomenological models that describe the rotational 
and vibrational bands \cite{rad10}. In honnour of his 70th birthday I dedicate the present work to him.

Throughout the Nineties, several new features of the $^{252}$Cf spontaneous
fission have been disclosed by means of the two-dimensional coincidence spectra analysis of the
prompt gamma rays emitted by fission fragments (FF) \cite{ham03} (Gammasphere).
Among these the determination of average angular momentum for primary fission fragments
as a function of neutron multiplicity for the Mo-Ba and Zr-Ce charge splits received due attention 
\cite{pop02}. 
The data indicates an increase in the fragment average spin up to $\nu=4$ emitted neutrons, followed by
a decrease down to values comparable to the cold fission case at $\nu=8-10$.  
Although the scenario proposed by the Gammasphere group in order to explain the population of rotational 
states in these neutron-rich fragments is not free of logical flaws, they did an essential observation :
there is an apparent correlation between the fragment angular momentum and the fragments elongation 
for increasing number of emitted neutrons. This conclusion was stems from the dependence of the measured 
FF average spin $\langle J_{\rm L,H}\rangle$ on the total number of emitted neutrons $\nu_{\rm tot}$. 
In a simple scheme of logical deduction, it can be admitted that with increasing $\nu_{\rm tot}$ the 
system gets more and more excited and consequently the energy stored into deformation increases. 
Consequently the FF have larger elongation, thus larger moments of inertia, and they correspondingly
display and enhanced propensity to rotation. When $\nu_{\rm tot}$ increases further, the spin carried away 
by the emitted neutrons from the fissioning system tends to diminish the average value of angular 
momentum in each fragment. The apparent symmetric shape of the function 
$\langle J_{\rm L,H}\rangle (\nu_{\rm tot})$ (see Fig.1) could indicate a smooth transition from the
ground-state deformed to well-deformed shapes of the FF up to a maximum followed by a transition back to sphericity 
with increasing values of the argument.

\begin{figure}
\centering    
\includegraphics[width=0.49\textwidth]{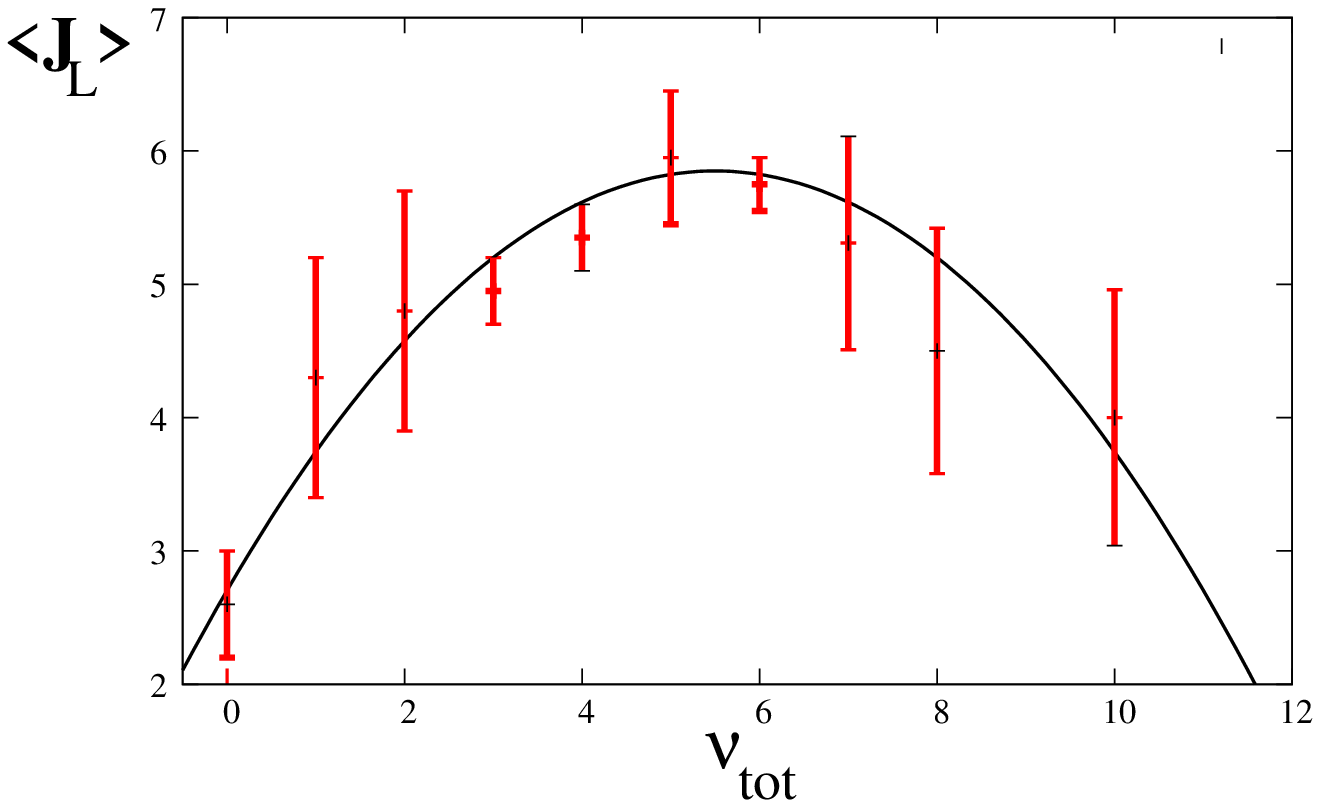}
\includegraphics[width=0.49\textwidth]{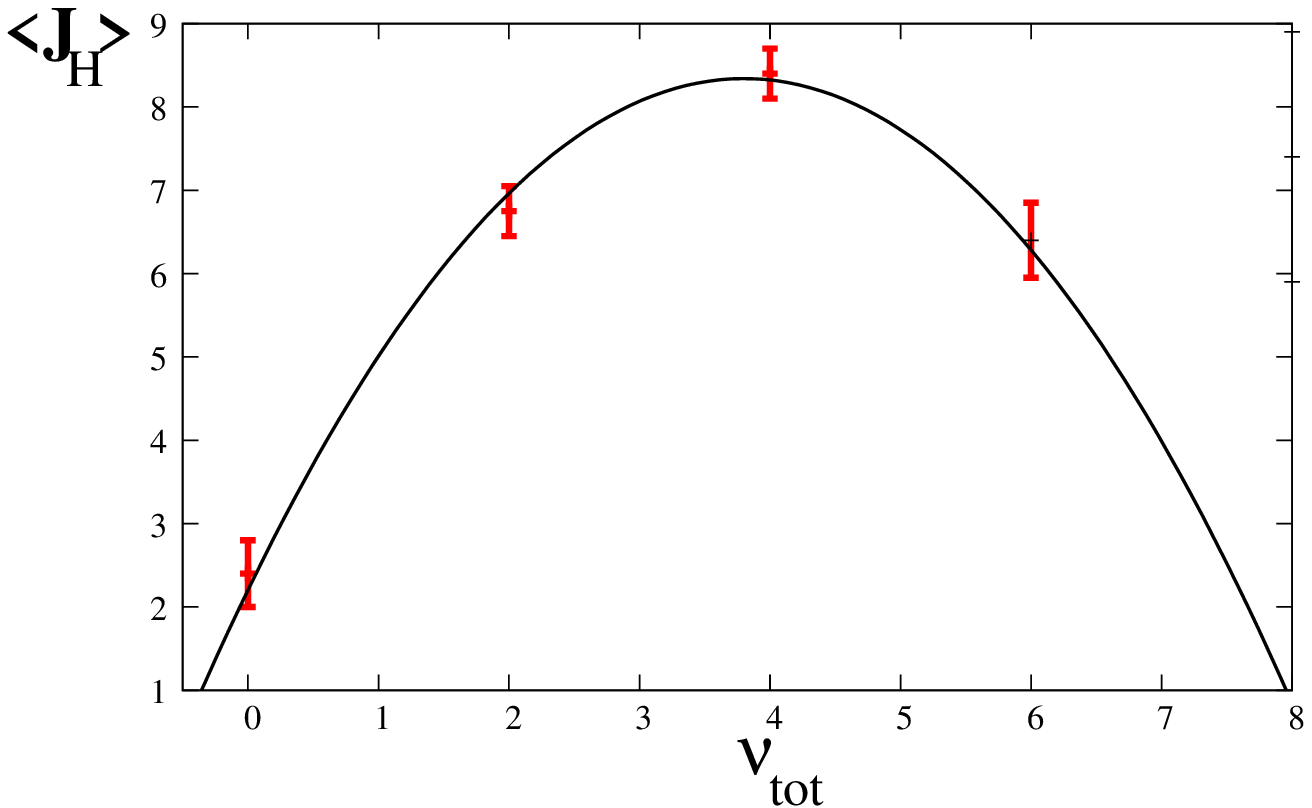}
\caption{Average spins of FF for the Mo-Ba splitting up to $\nu=8$ as reported in \cite{pop02}.}
\label{fig0}
\end{figure}

In a joint effort with collaborators from Bucharest and Frankfurt, we addressed the special case of cold 
(neutronless) fission in view of its resemblance to cluster radioactivity
\cite{san99}.  In ref.\cite{mis99} we assumed a scenario where the bending oscillations alone are responsible 
for angular momentum generation in fragments. Within the bending model the final angular momentum of each 
fragment is built up at scission and its generation is due solely to quantum fluctuations of the fragments 
orientation with respect to the strongly polarized fission axis.
In this picture, angular momentum of FF carry unaltered fingerprints of the 
scission configurations (deformation of fragments and distribution of intrinsic
excitatations among the fragments).  

\section{Spins of FF at scission}
 
By scission point we understand in what follows a configuration where the fragments, while still in touch, 
have well-defined shapes, whether they are formed in their ground states like in cold fission or are elongated 
in the direction of the fission axis according to the way excitation energy is shared among fragments.
In other words we adopt a cluster-like approach to fission as we did previously for cold fission \cite{mis99}.
In order to study the population of rotational states in the fission partners we have to take into account the 
deformation of nuclei and therefore establish the shape of the heavy-ion potential in the "orientation landscape", 
a task that we accomplish by appealing to the double-folding method. In this framework the separation of the radial 
from the angular coordinates in the potential is carried out by means of Fourier techniques as explained in
detail in our earlier work \cite{mis02} .

The calculation of the deformation energy at scission and the distribution 
of excitation energy among fragments is done by means of a general recipe which consists in determining first 
the ratio of prompt neutron numbers 
emitted by complementary fragments and then equate it to the ratio of fission fragment excitation energies. 
We employ a very recent analysis of the systematic behavior of experimental 
$\nu=\nu_L+\nu_H$ ($\nu_H/\nu$ as a function of $A_H$) \cite{man11}. Once $\nu_L$ and  $\nu_H$ are determined
the following relations between the excitation energy of complementary, fully accelerated fragments, 
and the total excitation energy $TXE$ are used
\beq
E_{H,L}^*=\frac{\nu_{H,L}}{\nu}TXE
\eeq
We assume that at scission the excitation is accounted by the energy stored in the fragments deformation 
and a proportionally smaller contribution spent on the excitation of intrinsisc degrees of freedom
\beq
TXE=E_{\rm def}+E_{\rm sc}^*
\eeq
The deformation energy is taken as the sum of liquid drop and shell model corrections energy
\beq
E_{\rm def}=\sum_{i=L,H}(E_{\rm LDM}^i+\delta W_{\rm shell}^i)
\eeq 
The LDM part is taken as the sum of a surface and Coulomb contributions only (the volume energy is taken to be 
independent of deformations) for a sharp-edge distribution of nuclear matter 
relative to the value for a sphere.
To estimate it we appeal to the embedded spheroid model \cite{bros90} . 
As for the shell corrections we apply the expeditive procedure of Myers and Swiatecki
\cite{myswi67} .

The content of $E_{\rm sc}^*$ is assumed to be accounted soleley by the excitation of 
collective rotational states of the FF. This quantity amounts to a few MeV regardless of 
the total number of emitted neutrons.

Following the spirit of the bending model \cite{mis99}, we assume that at scission the elongation (translational) 
degree of freedom $R=R_{\rm sc}$ is {\it nailed down}, and the FF can execute only angular vibrations 
around a direction perpendicular to the fission axis. Consequently the relevant degrees of freedom are the 
angular deviations 
$\theta_{1(2)}$ of the FF from the fission axis. The quantized bending Hamiltonian can be read off from eq.(18), 
ref.\cite{mis99} 
\beq
H_{\rm bend} = ~-\frac{\hbar^2}{2{\cal B}_1}\frac{\partial^2}{\partial\theta_1^2}
    ~-\frac{\hbar^2}{2{\cal B}_2}\frac{\partial^2}{\partial\theta_2^2}
    ~-\frac{\hbar^2}{\mu R_{\rm sc}^2}\frac{\partial^2}{\partial\theta_1
      \partial\theta_2}~+~W_{\rm int}(\theta_1,\theta_2)
\label{hamilt1}
\eeq
where ${\cal B}_{1(2)}$ are related to the inertia moments of the
fragments, ${\cal J}_{1(2)}$
\beq
{\cal B}_{1(2)}=\frac{{\cal J}_{1(2)}\mu R_{\rm sc}^2}{{\cal J}_{1(2)}+
\mu R_{\rm sc}^2}.
\eeq
In the Hamiltonian (\ref{hamilt1}) we take into account only the deformed part of the potential, 
i.e. we discard the monopole term
\beq
W_{\rm int}(\theta_1,\theta_2)=
{1\over 2}C_1\theta_1^2 + {1\over 2}C_2\theta_2^2
+C_{12}\theta_1\theta_2
\label{potcupl}
\eeq  
which is obtained by expanding (\ref{wint}) up to quadratic terms 
in powers of the polar angle $\theta.$
The explicit form of the stiffness parameters is \cite{mis99}. 
\beqa
C_{1,2} & = & -{1\over 2}\sum_{\lam_1\lam_2\lam_3}~\lam_{1,2}(\lam_{1,2}~+1)
V_{\lam_1\lam_2\lam_3}^{0~0~0}(R_{\rm sc})\\
C_{12} & = & - {1\over 4}\sum_{\lam_1\lam_2\lam_3}
\left \{ \lam_3(\lam_3~+1)  - \lam_1(\lam_1~+1) - \lam_2(\lam_2~+1)\right \}
V_{\lam_1\lam_2\lam_3}^{0~0~0}(R_{\rm sc})
\eeqa

The bending Hamiltonian, consisting of two coupled Hamiltonians, can be reduced to the canonical form 
by means of a unitary transformation of exponent
\beq
S = \theta_1\frac{\partial}{\partial\theta_2}  
-\frac{K_p\omega_2^2+K_q}{K_p\omega_1^2+K_q}
\theta_2\frac{\partial}{\partial\theta_1}
\eeq
where  
\beq
\omega_{1(2)} = \sqrt{C_{1(2)}\over {\cal B}_{1(2)}},~~~~~~
K_q = \frac{C_{12}}{\sqrt{{\cal B}_1{\cal B}_2}},~~~~~~~
K_p=\frac{\sqrt{{\cal B}_1{\cal B}_2}}{\mu R_{\rm sc}^2}.
\eeq

The eigenvalues of the uncoupled Hamiltonian have the usual form
\beq
E_{n_1n_2} = \hbar\Omega_1\left ( n_1+{1\over 2}\right ) +
             \hbar\Omega_2\left ( n_2+{1\over 2}\right )
\label{energben}
\eeq
where the relation between the new frequencies $\Omega_{i}$ and the
old ones $\omega_i$, if we choose $\omega_{1} > \omega_{2}$, is given by
\beq
\Omega_{1(2)}^2  =
\frac{\left ( K_q(\omega_1^2+\omega_2^2\pm\sqrt{\Delta})
+2K_p\omega_1^2\omega_2^2\right)
\left( K_p(\omega_1^2+\omega_2^2\pm\sqrt{\Delta})
+2K_q\right )}
{4(K_p\omega_1^2+K_q)(K_p\omega_2^2+K_q)}
\eeq
with
\beq
\Delta = (\omega_1^2-\omega_2^2)^2 +
4(K_p\omega_1^2+K_q)(K_p\omega_2^2+K_q)
\eeq
In the bending picture the angular momentum of each fragment is calculated as square-roots of 
the ground state ($(n_1,n_2)=(0,0)$) expectation values of the operators 
$L_k^2=-\hbar^2\partial^2/\partial\theta_k^2(k=1,2)$ \cite{mis99}.

\section{Evolution of FF spins during the post-scission stage}

In the post-scission stage, the decay of $^{252}$Cf in two fragments is described by the  
the distance between the fragments centers $R$ which stands for the 
elongation of the system undergoing fission and 
by he orientation in space $\Omega_i$
of each rotator.

Fragment deformations ($\beta_i$) are considered as parameters of the  
dynamical problem and they preserve the value calculated at scission for 
a given excitation energy.    
In what follows we are going to specify the fragment-to-fragment distance $R$
as a variable depending parametrically on time, i.e. $R=R(t)$. 

Accordingly, the dynamical problem is formulated in terms of a Hamiltonian where the translational 
and rotational motion  are coupled only via the deformed part of the potential.
\beq
H_{\rm fiss}=\frac{1}{2}{\mu\dot{R}^2}
+V_0(R)+H_{\rm rot}(\Omega_{1})+H_{\rm rot}(\Omega_{2})+W_{\rm int}(R,\Omega_1,\Omega_2)
\eeq
In the above formula the first term represents the translational kinetic energy that 
asymptotically coincides with the observed kinetic energy $TKE(\infty)$, i.e.
\beq
TKE=Q-TXE
\eeq
whereas 
\beq
H_{\rm rot}(\Omega_{1,2})=\frac{\bd{L}_1^2}{2 {\cal J}_{\rm 1,2}}
\eeq
are the free rotational Hamiltonians of the fragments with angular momentum $\bd{L}_i$. 
The non-trivial part of our fission Hamiltonian is represented by the deformed part
of the interaction from which we substracted the monopole-monopole interaction $V_0$ 
(see \cite{mis02} for the derivation) :
\beq
W_{\rm int}(R(t),\Omega_1,\Omega_2)=\sum V_{\lambda_1\lambda_2\lambda_3}
^{\mu-\mu 0}(R(t))D_{\lambda_1 0}^\mu(\Omega_1)D_{\lambda_2 0}^{-\mu}(\Omega_2)-V_0(R(t))
\label{wint}
\eeq
Whereas the standard approach \cite{eisgre70} to Coulomb excitation (Coulex) assumes that the relative motion
of the centers of mass of the two reacting nuclei can be described clasically, in the present
approach we calculate the time-evolution of a quantum wave-packet across the Coulomb+nucler barrier.
Note that in the case of Coulex the classical assumption for the trajectory $R(t)$ is
justified by large impact parameters. This has the obvious consequence that the two nuclei 
do not get into the domain of nuclear interation and therefore the quantum tunneling across
the reciprocal fusion barrier plays no role. For that reason we stick to the other approximation used in 
the Coulex case, namely that
the energy transfer between the two nuclei is negligible. This assumption is likely to be valid
in fission, since it seems to be not unreasonable that after scission
the fragments are only weakly redistributing the excitation energy stored in intrinsic motion;
excitation energy stored in deformation is shared between fragments at scisssion and do not 
get changed afterwards as we alredy remarked above.
Since the translational energy exceeds by far the energy distributed on the excited rotational 
states (approximately a factor of 100), it is justified to assume that quantum tunneling  
is the dominant part of the Hamiltonian and its dynamics can be resolved separately from the
one corresponding to the rotational degrees of freedom.

In support of the above assumption we should also mention that due to the barrier thickness, 
at least in the quantum tunneling regime, $R(t)$ represents the slow degree of freedom. 
The period of bending oscillations is much shorter than the 
time necessary for the wave packet average position to cover the distance between the first and the 
second turning point.  

Under such circumstances the relative motion of the two flying apart fragments is not affected 
by the excitation of rotational levels in the two rotating fragments and therefore the total 
wave-function is given by the direct product of radial and rotational wave functions:
\beq
\Psi\sim \psi(R,t)\otimes \Phi(\Omega_1,\Omega_2;R(t))
\label{factwf}
\eeq 
Note that in the rotational wave-function, $R(t)$ appears as a parameter. 

The relative motion is governed by the time-dependent Schr\"odinger equation 
in this degree of freedom 
\beq
i\hbar\frac{\partial\psi(R,t)}{\partial t} =
\left \{ -\frac{\hbar^2}{2\mu}\nabla_{\bd{R}} + V_0(\bd{R})\right \}\psi(R,t)
\label{tdse}
\eeq
The initial wave-packet is prepared as a bound state of energy $TKE$, 
in the spherical potential $V_0(R)$ for a two-body system of reduced mass $\mu$, 
i.e.
\beq
-\frac{\hbar^2}{2\mu}\Delta_{\bd{R}}\psi(R,0)+V_0(R)\psi(R,0)=TKE\psi(R,0)
\eeq  
and next it is propagated in time by applying the Crank-Nicolson scheme  as we did in previous
publications on $\alpha$-decay \cite{misriz00} and cold fission \cite{missan01}. 
The quantum trajectory $R(t)$ can now be extracted from the knwoledge of the wave-packet 
$\psi(R,t)$ at any moment of time,
\beq
R(t) = \int_0^{\infty}d~R~R~|\psi(R,t)|^2
\eeq
Note that the evolution of the wave-packet is determined for a time long enough to include 
not only the sub-barrier, but also the pure Coulomb part of the trajectory.  

In view of the factorization (\ref{factwf}) the time-dependent Schr\"odinger equation 
describing the coupled dynamics of the two rotators reads
\beq
i\hbar\frac{\partial\Phi(t)}{\partial t} =
\left [ H_0+W_{\rm int}(t)\right ]\Phi(t)
\label{tdserot}
\eeq
where $H_0$ is the Hamiltonian of the free (uncoupled) rotators wich satisfying the
eigenvalue problem
\beq
H_0\mid (I_1M_1)(I_2 M_2) \rangle = E_{I_1I_2}\mid (I_1M_1)(I_2 M_2) \rangle 
\eeq 
$I$ and $M$ are the spin and the magnetic quantum number respectively.
In what follows we resort to the approximation $M_1=M_2=0$, used also in the framework of the
bending model, and justified by the strong polarization of fragments at the scission moment ($t=0$).
In this configuration the spin of each fragment is oriented along an axis perpendicular to the
fission axis \cite{mis02} and its projection on the fragment $z$-axis is nil.    

The matrix elements of the interaction in the basis $\mid I_1 I_2\rangle$ are readily calculated 
using standard angular algebra techniques
\beqa
\langle I_1 I_2 \mid W_{\rm int}\mid I^\prime_1 I^\prime_2\rangle&=&
\frac{1}{4\pi}\frac{{\hat I}_1{\hat I}_2}{{\hat I^\prime_1}{\hat I^\prime_2}}\times\nonumber\\
&&\sum_{\lambda_1+\lambda_2+\lambda_3\geq 4} {\hat \lambda_1} {\hat \lambda_2}
{V}_{~\lam_1\lam_2\lam_3}^{~0~0~0}(R)\left (C^{I_1\lam_1I^\prime_1}_{0~0~0}\right )^2
\left (C^{I_2\lam_2I^\prime_2}_{0~0~0}\right )^2\nonumber
\eeqa

\begin{figure}
\centering
\includegraphics[width=1.1\textwidth]{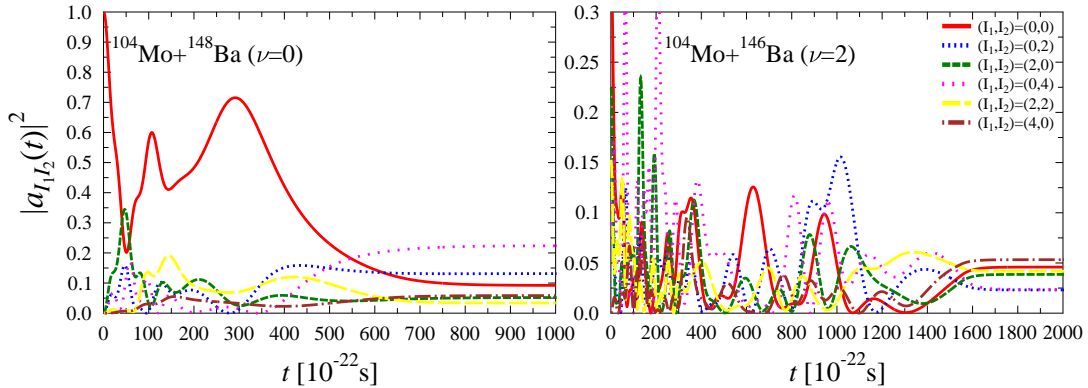}
\caption{Evolution of the modules of the rotational states amplitudes.}
\label{fig1}
\end{figure}

To solve eq.(\ref{tdserot}), $\Phi(t)$ is expanded in eigenstates of $H_0$, i.e.
\beq
\Phi(t)=\sum_{I_1I_2}{a}_{I_1I_2}(t) e^{ -\frac{i}{\hbar}E_{I_1I_2}t}\mid I_10\rangle \mid I_20\rangle
\eeq  
and thus instead of solving a two-dimensional parabolic equation with a time-dependent potential,
that provides a solution that has to be subsequently projected on each eigenstate 
$\mid I_1 I_2\rangle$, we arrive at a coupled system of linear ordinary differential equations 
\beq
i\hbar {\dot a}_{I_1I_2}(t)=\sum_{I^\prime_1 I^\prime_2}\langle I_1 I_2 \mid W_{\rm int}\mid I^\prime_1 I^\prime_2\rangle
\exp{\left [ i(E_{I_1I_2}- E_{I^\prime_1I^\prime_2})t/\hbar\right ]}
a_{I^\prime_1 I^\prime_2}
\eeq   
The initial values $a_{I_1I_2}(0)$ are fixed by the spins evaluated at the scission point.    
In the basis we include up to 36 states, which means that in each fragment a state of maximum spin $I=14$
can  be populated. Separating the real and imaginary components of the complex amplitude 
$a_{I^\prime_1 I^\prime_2}$ we end-up with a system of 72 equations. Even in the case when we include 
multipoles  up to $\lambda=6$ in the potential, a run on a standard computer takes less time than 
solving the partial differential equations (\ref{tdserot}).
The amplitudes above depend on the quantum trajectory $R(t)$, which for thick barriers has a strongly oscillating
behavior in the tunneling regime and then, outside the Coulomb barrier, tends asymptotically to the 
classical trajectory. 
In Fig.\ref{fig1} we plot the time evolution of the squared modulus amplitudes $\mid a_{I_1I_2}(t)\mid^2$.
On the left panel we represented the neutronless ($\nu=0$) case. Since no energy is available for rotation at
scission, the initial probability is concentrated in $a_{00}$ and therefore the rotational g.s. state $|00\rangle$ 
dominates mostly during the sub-barrier motion. When fragments are running in the pure Coulomb field, it will be 
overunned by higher spin states of the heavier fragment, e.g. $|04\rangle$ and     $|02\rangle$.
In the case with $\nu=2$ (right panel of Fig.\ref{fig1}), the non-rotating state  $|00\rangle$ decrease rapidly 
during the sub-barrier motion and displays some revivals (flashes) with smaller amplitude as time goes on.
One should note that in the case of Coulomb excitation, when one deals with smooth non-oscillatory 
trajectories, there is a complete revival of the amplitudes \cite{fonda88} . 
In this case the population probability is higher 
in the light fragment, the states $|40\rangle$, $|20\rangle$ prevailing over $|04\rangle$ and $|02\rangle$

\begin{figure}
\centering    
\includegraphics[width=1.1\textwidth]{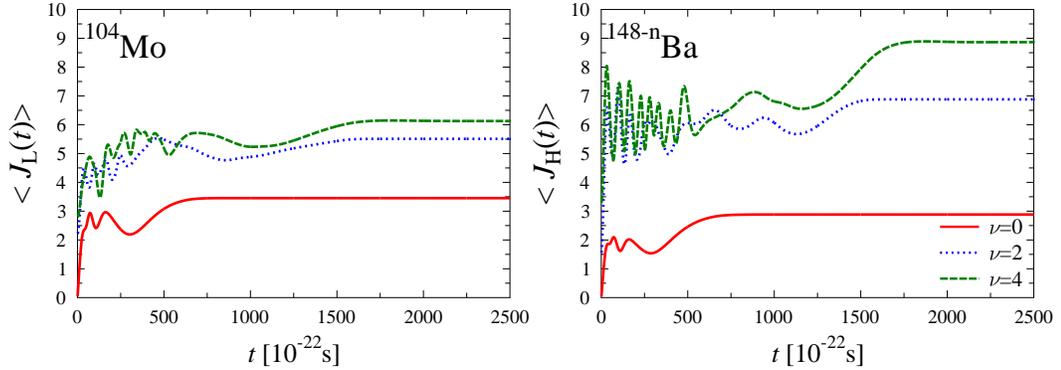}
\caption{Evolution of average spins of the light (Mo) and heavy (Ba) fragments for total number
of emitted neutrons $\nu=0,2,4$.}
\label{fig2}
\end{figure}

The time-dependent average values of the spins are simply given in terms of the amplitudes 
 \beq
\langle J_{\rm L,H}^2\rangle =\sum_{I_1 I_2}\mid a_{I_1 I_2}(t)\mid^2 I_{1,2}(I_{1,2}+1)
\eeq
In Fig.\ref{fig2} the time evolution of the average momentum is displayed separately for the light 
fragment (left panel) and the heavy fragment (right panel) for 3 different excitation energies that
correspond to $\nu=0,2,4$. We should note at this point that in our calculations we do not take into
account any type of correction of the average momenta of the primary fragments for the loss from neutron 
evaporation.
The data reported in \cite{pop02} operates such a correction by using an evaporation calculation for the spin 
$j$ removed on average by one evaporated neutron. In that reference they reported calculated $j$ values 
for Ba and Mo varrying from
$\sim 0.6\hbar$ to $\sim 0.3\hbar$  when $\nu_L$ changed from 1 to 6 and $\nu_H$ from 1 to 3.
Even if we adopt the most simple assumption, i.e. each removed neutron carries away $j=1/2$, the fact that the 
average angular momentum increases for both fragments still hold according to Fig.\ref{fig2}.

\section{Conclusions}

The fission dynamical model used in this study with the aim to estimate the average of spins of 
the emitted fragments consists of a traditional statical approach at scission, where the initial  
population of rotational is determined and subsequently is feeding a semiclasscial time-dependent coupled-system
of equations for the occupation amplitudes of these states. 
The non-stationary quantum sub-barrier motion induces a strongly non-linear behavior of these 
amplitudes which eventually stabilizes to constant asymptotic values in the purely Coulomb field.
Consequently the generation of angular momentum is the result of the complex dynamics 
after scission. Previous attempts, based exclusively on static models, that addressed the quest of how rotation is
pumped into the intially "frozen" fragments are in our opinion limited to a relatively early instance 
of the fission process and therefore cannot provide credible predictions for the final characteristics of 
the FF. 
   
The model propsed in this work confirms the observed steady increase of the fragments 
rotation in spontaneous fission of  $^{252}$Cf when going from 1 to 4 
emitted neutrons. However, knowledge on the fragment configuration and energy content at scission can 
only indirectly be extracted from such an analysis. 

The present approache is able to provide estimations also for other observables in binary fission such as 
the translational kinetic energy and can be extended to ternary fission.  


\section*{Acknowledgement}
This work received financial support from UEFISCDI Romania under the programme
PN-II contract no. 116/05.10.2011.
I am gratefull to Dipl.Phys. C. Matei for assistance with the 
artwork. To Dr.M. Jandel and Prof.J. Kliman I am indebted for enlightning discussions 
regarding the experimental state-of-art.


\end{document}